\documentclass[hyper]{JINST}

\pdfoutput=1

\usepackage{booktabs}
\usepackage{lineno}
\usepackage{mhchem}
\usepackage{upgreek}
\usepackage{wasysym}
\usepackage{subfig}
\usepackage{multirow}

\newcommand{\gerda}{\textsc{Gerda}}
\newcommand{\onbb}{$0\nu\beta\beta$}
\newcommand{\Ge}{$^{76}$Ge}
\newcommand{\Qbb}{Q$_{\beta\beta}$}

\newcommand{\eventsper}{events$/($keV$\cdot$kg$\cdot$yr$)$}

\newcommand{\eventsperkBq}{events$/($keV$\cdot$kg$\cdot$yr$\cdot$kBq$)$}
\newcommand{\Th}{$^{228}$Th}

\title{Production and Characterization of $^{\textbf{228}}$Th Calibration Sources with Low Neutron Emission for GERDA}

\author{L.~Baudis$^a$,
  G.~Benato$^a$\thanks{Corresponding author.}\ ,
  P.~Carconi$^b$,
  C.~Cattadori$^c$,
  P.~De~Felice$^b$,
  K.~Eberhardt$^d$,
  R.~Eichler$^e$,
  A.~Petrucci$^b$,
  M.~Tarka$^a$\thanks{Now at Department of Physics and Astronomy, Stony Brook University, Stony Brook, NY 11794-3800, USA}\ \ 
  and M.~Walter$^a$\\
  \llap{$^a$}Physik Institut der Universit\"at Z\"urich,\\
  Winterthurerstrasse 190, CH-8057 Z\"urich, Switzerland\\
  \llap{$^b$}ENEA, Istituto Nazionale di Metrologia delle Radiazioni Ionizzanti (INMRI),\\
  Centro Ricerche Casaccia, Via Anguillarese 30, I-00123 Rome, Italy\\
  \llap{$^c$}INFN Milano Bicocca,\\
  Piazza della Scienza 3, I-20126 Milano, Italy\\
  \llap{$^d$}Institut f\"ur Kernchemie, Universit\"at Mainz,\\
  Fritz-Strassmann-Weg 2, D-55128 Mainz, Germany\\
  \llap{$^e$}Paul Scherrer Institut,\\
  CH-5232 Villigen, Switzerland\\

  E-mail: \email{gbenato@physik.uzh.ch}}

\abstract{The \gerda\ experiment at the Laboratori Nazionali del Gran Sasso (LNGS)
  searches for the neutrinoless double beta decay of $^{76}$Ge.
  In view of the \gerda\ Phase~II data collection,
  four new \Th\ radioactive sources for the calibration of the germanium detectors
  enriched in $^{76}$Ge have been produced with a new technique,
  leading to a reduced neutron flux from $(\alpha,n)$\ reactions.

  The gamma activities of the sources were determined with a total uncertainty of $\sim4\%$\
  using an ultra-low background HPGe detector operated underground at LNGS.
  The emitted neutron flux was determined using a low background LiI(Eu) detector
  and a $^3$He counter at LNGS. In both cases, the measured neutron activity is $\sim10^{-6}$~n/(s$\cdot$Bq),
  with a reduction of about one order of magnitude with respect to commercially available \Th\ sources.
  Additionally, a specific leak test with a sensitivity to leaks down to $\sim10$~mBq was developed
  to investigate the tightness of the stainless steel capsules housing the sources after their use in cryogenic environment.}


\begin{document}

\section{Introduction}\label{sec:introduction}

The GERmanium Detector Array (\gerda)~\cite{gerdainstrument} is an experiment
searching for neutrinoless double beta decay (\onbb) in \Ge\
at the Gran Sasso National Laboratory of INFN (Italy).
In \gerda, the germanium crystals, enriched to $\sim86\%$\ in \Ge,
are simultaneously operated as source and detectors.
They are immersed in $64$~m$^3$\ of liquid argon (LAr), acting as cooling medium
and shield against external radiation.
The expected \onbb\ decay signature is a peak at the Q-value of the reaction (\Qbb), 2039~keV for \Ge~\cite{douysset}.
The \gerda\ physics program is divided in two stages: the first physics run, denoted as Phase~I,
took place between November 2011 and June 2013 with about 15~kg of \Ge\
and a background index (BI) at \Qbb\ of $10^{\mbox{-}2}$~\eventsper.
In the second data collection period, Phase~II, the active mass will be doubled
and a BI of $10^{\mbox{-}3}$~\eventsper\ is aimed for~\cite{procTIPP}.
This is achievable with the use of materials with higher radio-purity,
the application of pulse shape discrimination (PSD)
and the implementation of an active anti-coincidence veto in LAr.

The energy calibration of the germanium detectors is performed on a weekly basis.
Up to three calibration sources can be lowered in LAr in the vicinity of the detectors
through a dedicated calibration system~\cite{tarka,francispaper,francisthesis} developed and built at the University of Zurich.
In order to avoid radon contamination of LAr,
the sources cannot be removed from the experimental setup during physics runs.
They are therefore parked in gaseous argon on the top of the LAr cryostat, at the maximum possible distance from the detectors.
In addition, a tungsten absorber is mounted below the sources to further shield the detectors from gamma radiation.

The choice of the isotope for the sources is based on several requirements.
First, a substantial number of gamma lines has to be available for the calibration
of the energy scale up to \Qbb. Second, the half life has to be at least of the same order
of the experiment's live time, i.e. a few years. Finally, the PSD methods
for germanium detectors are usually tuned with double escape events,
which have very similar signal properties to the expected \onbb\ decay events~\cite{gerdapsd}.
For this reason, a double escape peak (DEP) with sufficient statistic has to be available.
The candidate which optimally fulfills these requirements is \Th,
with a 1.9~yr half life, a dozen of high statistic lines between 500~keV and 2.6~MeV and a DEP at 1592.5~keV~\cite{francisthesis}.

The main drawback of using \Th\ sources is that, starting from a \Th\ nucleus,
5 $\alpha$~particles with energies between 5.2 and 8.79~MeV are emitted
before a stable nucleus ($^{208}$Pb) is reached~\cite{nudat2}.
If the radioactive substance is embedded in materials with low threshold for $(\alpha,n)$~reactions,
a parasitic neutron flux is obtained. This is the case for standard commercial sources,
for which ceramic components, e.g. \ce{NaAlSiO2}, are used for practical reasons.
In \gerda, neutrons could generate $^{77}$Ge and $^{77m}$Ge via \Ge\ activation,
both of which undergo beta decay with $>2$~MeV Q-value, or they could be captured in the materials
surrounding the detectors and produce high energy gamma rays.
The background contribution at \Qbb\ prior to the application of pulse shape discrimination
and liquid argon veto cuts induced by such neutrons would be
$3\cdot10^{-5}$~\eventsperkBq~\cite{tarka}. With a total activity of $\sim 70$~kBq,
this would not fulfill the background requirements of \gerda\ Phase~II~\cite{procTIPP}.
The reduction of the neutron source strength by about one order of magnitude
can be achieved by embedding \Th\ in a metallic material with cross section for $(\alpha,n)$\ reactions
higher than 8.8~MeV. As described in~\cite{tarka,werner},
gold is the best candidate due to its 9.94~MeV threshold, its $<20~\upmu$m range for \Th\ alphas,
its ductility and ease of procurement.
This paper describes the production and characterization in terms of gamma and neutron activity
of four custom made \Th\ sources for \gerda\ Phase~II.

Given the ultra-low background requirements of the \gerda\ experiment
and the need to deploy the sources at cryogenic temperatures,
it is crucial to avoid any leakage of the source active material.
Thus, a new leak test procedure with a sensitivity to leaks of $\lesssim10$~mBq activity was developed
at the ENEA-INMRI, Italy, and tested for the first time on the \gerda\ Phase~II sources.

In Section~\ref{sec:sourceproduction} the production of low-neutron emission \Th~sources is revised.
Sections~\ref{sec:sourceactivity} and~\ref{sec:neutronstrength} describe the measurements
and the estimation of the sources activity and neutron strength.
In Section~\ref{sec:leaktest} the results of leak test after use in cryogenic environment are reported.

 \section{Custom production of $^{\mathbf{228}}$Th calibration sources}\label{sec:sourceproduction}

The production technique of \Th\ sources with reduced neutron emission was originally developed
by the University of Zurich (UZH) and the Paul Scherrer Institute (PSI).
Several prototype sources were produced during \gerda\ Phase~I~\cite{tarka}.
The technique was later used and improved~\cite{werner} to produce a strong calibration source of several MBq activity for Borexino.
A brief summary of the technique is given below.

The radioactive substance was provided in form of \ce{ThCl4} dissolved in 1~M \ce{HCl} by Eckert und Ziegler
Isotope Products Int., Valencia (CA), USA, with a total \Th\ activity of $150\pm23$~kBq.
Chlorine has two stable isotopes, $^{35}$Cl and $^{37}$Cl, both of which have $(\alpha,n)$\ thresholds below 8.8~MeV.
It is therefore mandatory to completely separate chlorine from thorium prior to deposition on gold.
The thorium tetrachloride solution was evaporated in a PTFE crucible almost to dryness;
it was subsequently worked-up two times with 1 ml concentrated nitric acid and dried by evaporation.
Thus the \ce{ThCl4} was converted in \ce{Th(NO3)4} using surplus concentrated \ce{HNO3} through the reaction:
\begin{equation}
\ce{3ThCl4 + 16HNO3 ->[\sim 115^{\circ}C] [3Th(NO3)4]_s + [4NOCl + 8H2O + 4Cl2]_g}
\end{equation}
where the subscripts $s$\ and $g$\ stand for solid and gaseous, respectively.
Subsequently, the solid \ce{Th(NO3)4} was diluted in two molar \ce{HNO3} and transferred
into a gold crucible prepared out of a  $2\times2$~cm gold foil of $20~\upmu$m thickness.
The gold has a $\geq99.99\%$\ purity and was produced by mechanical means,
lamination and hammering, with no alteration of the original purity level~\cite{battiloro}.
The \ce{HNO3} was evaporated by heating the solution to $\sim120^{\circ}$C.
Subsequently, the gold crucible was folded and heated to $\sim700^{\circ}$C to form \ce{ThO2}
on the gold surface to evaporate the remaining nitrogen and oxygen through the reaction:
\begin{equation}
\ce{3Th(NO3)4 ->[\sim700^{\circ}C] [3ThO2]_s + [12NO2 + 3O2]_g }
\end{equation}
Finally, the gold crucible was further folded and wrapped in a second gold foil of the same dimension
in order to prevent any loss of radioactive material.
The procedure was performed separately for each of the four sources.
The production of the \gerda\ Phase~II sources
was performed at the Institute for Nuclear Chemistry of the University of Mainz, Germany.

The sources were then encapsulated by Eckert und Ziegler Nuclitec GmbH, Braunschweig, Germany,
with a double-sealed stainless steel VZ-3474 capsule, and certified according to ISO-2919 requirements.

\section{Measurement of source activity}\label{sec:sourceactivity}

A precise estimation of the sources activity is required for several reasons.
First, a correct \onbb\ decay analysis can be performed only if
a proper understanding of all the background components is available.
This is possible thanks to the development of a background model via Monte Carlo (MC) approach
and the fit of the model to the physics data.
The correctness of the MC geometry and physics implementation can be assessed
by simulating the calibration measurements and comparing the simulated with the experimental spectra
recorded by the germanium detectors.
In \gerda\ Phase~I, the activity of the calibration sources was known with a $15\%$\ uncertainty~\cite{tarka},
hence the quality of the MC model could only be inferred by comparing the spectral shape
of simulated and measured spectra, and not by the absolute number of events.
Second, the simulation of the calibration measurement is required for the estimation
of the PSD efficiency on the DEP, the Compton continuum and the full energy peaks~\cite{gerdapsd}.
Finally, an estimation of the LAr veto efficiency for different background source locations
can be addressed by lowering the \Th\ sources to different positions
and comparing the measured with the simulated suppression factors.

The measurement of the Phase~II \Th\ source activities was performed in the Gator facility~\cite{gator} at LNGS.
Gator consists of a high-purity p-type coaxial germanium crystal encased in an ultra-low activity copper cryostat.
The detector is placed in a $25\times25\times33$~cm$^3$\ cavity where the material samples can be inserted.
The integrated background of Gator is $\sim0.16$~events$/$min in the 100-2700~keV range.
The $\sim30$~kBq activity of Phase~II sources induces a count rate which is 4 orders of magnitude higher,
hence the measurements described here can be considered background free.

A 20~min long measurement was performed for each of the four sources,
placed on a PTFE holder at $\sim120$~mm distance from the top of the detector to reduce the pile-up rate.
The dead time was between 16 and $25\%$, depending on the source activity.
The spectrum is recorded in the 10-2770~keV range,
thus covering the entire \Th\ spectrum, with the highest energy line at 2614.5~keV.

In order to estimate the activity, a {\sc{Geant4}} based simulation was performed~\cite{geant4}.
A negligible fraction of events are present in the experimental spectrum above the 2614.5~keV peak.
These are pile-up events and are not accounted for in the MC.
Similarly, a $\sim25\%$\  discrepancy between data and MC is visible below $\sim70$~keV.
This is attributed to a sub-optimal implementation of the \Th\ decay chain in {\sc{Geant4}}.
To avoid any bias due to the two effects explained above, the analysis is performed in the 100-2617~keV range.

For the comparison of measured and simulated data
the finite experimental energy resolution is applied to the simulated spectrum
via energy smearing on a single event basis.
The activity is estimated with a maximum likelihood analysis performed with BAT~\cite{bat}.
The log-likelihood is written as:
\begin{equation}\label{eq:likelihood}
\ln{\mathcal{L}} = \sum_{\text{bin}~i} \ln{ \Biggl( \frac{ \lambda_i^{k_i} \ \ e^{-\lambda_i}}{ k_i! } \Biggr) }
\end{equation}
where $k_i$\ is the measured number of events and $\lambda_i$\ the expectation value in the bin $i$, defined as:
\begin{equation}\label{eq:lambda}
\lambda_i = \frac{ A \cdot \Delta t \cdot R_i }{ N_{MC} }
\end{equation}
$A$\ is the source activity, $\Delta t$\ is the live time of the measurement,
$R_i$\ the number of events in the $i$-th bin of the simulated spectrum
and $N_{MC}$\ the total number of simulated events.
The only free parameters is $A$, its best value is given
by the maximum of the $\ln{\mathcal{L}}$, and its uncertainty is obtained
from the central $68\%$\ interval.

\begin{figure}[]
  \centering
  \def\svgwidth{\textwidth}
  \input{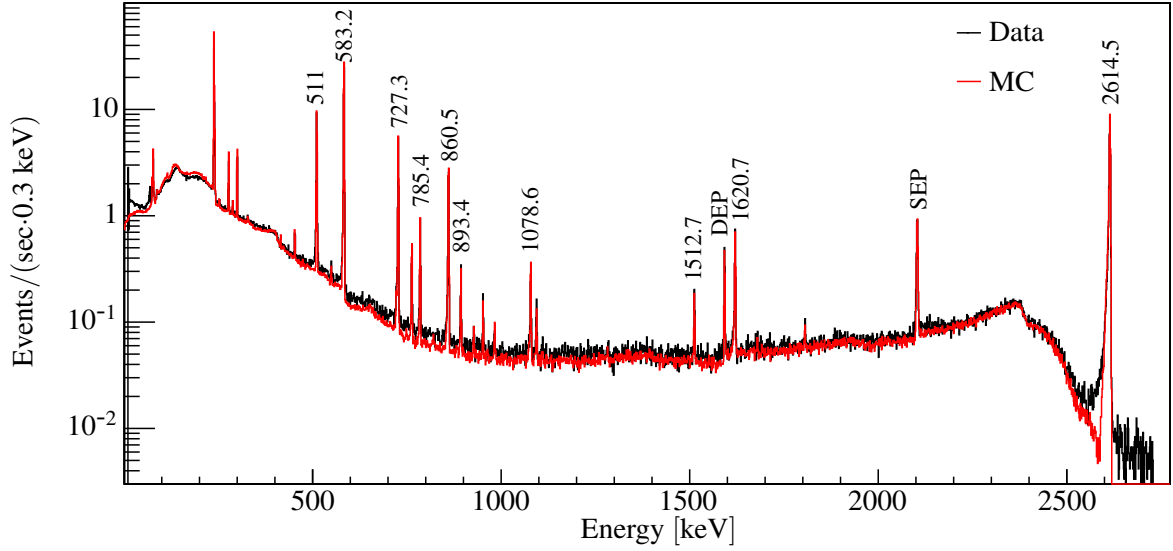}
  \caption{  \label{fig:spectrumDataMC}
    \Th\ spectrum recorded by the Gator spectrometer with the AD9857 source (black),
    together with the simulated spectrum (red) normalized according to the best fit.
    The gamma lines used for the energy calibration in \gerda\ are labeled.
    SEP and DEP stand for single and double escape peak, respectively.}
\end{figure}

The experimental spectrum, along with the simulated one scaled according to the best-fit result,
is shown in Fig.~\ref{fig:spectrumDataMC} for one of the sources.
The good agreement between data and MC at all energies in the considered range
together with the high statistics of the measurement are reflected in a statistical uncertainty
on the activity of $\sim2\permil$. The estimated activity of the four Phase~II sources
is reported in Tab.~\ref{tab:activity}.

\begin{table}
  \caption{Estimated activity of the four \gerda\ Phase II \Th\ calibration sources.
    The first error is to the statistical uncertainty of the fit,
    while the second is the total systematic uncertainty.}\label{tab:activity}
  \centering
  \begin{tabular}{cc}
    \toprule
    Source & Activity [kBq] \\
    \midrule
    AD9854 & $24.2\pm0.1$~(stat)~$^{+1.1}_{-0.9}$~(syst) \\
    AD9855 & $34.2\pm0.1$~(stat)~$^{+1.5}_{-1.3}$~(syst) \\
    AD9856 & $30.8\pm0.1$~(stat)~$^{+1.4}_{-1.2}$~(syst) \\
    AD9857 & $41.3\pm0.1$~(stat)~$^{+1.8}_{-1.6}$~(syst) \\
    \bottomrule
  \end{tabular}
\end{table}

The major source of uncertainty for this measurement is of systematic origin.
It is due to the limited knowledge of the experiment's geometry,
to an imprecise knowledge of the cross sections for the physics processes involved,
and to the fitting procedure itself.

Regarding the implementation of the geometry in the MC simulation,
the major uncertainty is induced by the distance between the source and the germanium crystal,
known with $\pm2$~mm precision.
All other uncertainties connected to the geometry and thickness
of the materials in between can be considered negligible.
A difference of $\pm2$~mm distance has negligible effect in the spectral shape but results
in a different covered solid angle and, therefore, a different activity.
The simulations and the analysis have been repeated after changing the source height by $\pm2$~mm.
This results in a $^{+2.5}_{-2.1}\%$~systematic error on the estimated activity.

The uncertainty due to the limited accuracy of the physics models implemented in {\sc{Geant4}}
is almost exclusively related to that of the cross sections for the photo-electric,
Compton and pair-production processes.
The accuracy of these was estimated to be at a $5\%$\ level, as described in~\cite{geant4validation}.
As for the previous case, the simulations have been repeated after separately changing the cross sections
of each of the three processes by $\pm5\%$.
Given the considered energy range, the largest effect on the activity is induced by the cross section for Compton scattering
and is $^{-2.0}_{+2.5}\%$. The systematic error related to the photo-electric effect is $^{-0.4}_{+0.2}\%$,
while the one related to the pair-production is below $1\permil$.

Finally, the uncertainty given by the choice of the energy range used for the analysis has to be considered.
To quantify this effect, the analysis has been repeated by increasing the minimum of the energy range from 100 to 2600~keV with 20~keV steps.
The root mean square (RMS) of the resulting activities distribution has been taken
as systematic error. This results in $2.5\%$.

The three sources of systematic uncertainties considered so far originate from independent causes.
Therefore, the errors are summed in quadrature, resulting in a $^{+4.4}_{-3.9}\%$\ uncertainty in the activity.
As a comparison, the activity of the \gerda\ Phase~I sources was only known with a $\pm15\%$~uncertainty~\cite{tarka}.
Given the high precision obtained in the measurement, a more precise validation of the \gerda\ Phase~II MC simulation is possible.
In addition, the measurement of $^{56}$Co and $^{226}$Ra radioactive sources for \gerda\ Phase~II will be performed
using the same procedure.

\section{Neutron strength measurement}\label{sec:neutronstrength}

The measurement of the neutron source strength of the four \gerda\ Phase II sources
was performed with a LiI(Eu) detector and a $^{3}$He counter
underground at LNGS in a low background environment.
As both detectors are suitable for the detection of thermal neutrons,
a moderator had to be used in order to thermalize the neutrons emitted by the \Th\ sources.
Three measurements were performed:
first, the total detection efficiency was determined using an AmBe neutron source of known neutron strength,
keeping the same geometrical configuration as for the \Th\ sources screening.
In a second step, the background spectrum was acquired for a period long enough
to make its influence on the neutron source strength uncertainty smaller than the systematic error.
Finally, the four \gerda\ Phase II sources were measured.

\subsection{Measurement with LiI(Eu) detector}\label{subsec:LiI}

The first measurement of the neutron source strength was performed with a LiI(Eu) detector.
The physical process exploited for the detection of thermal neutrons is:
\begin{equation}\label{eq:LiI}
^6_3\text{Li} + \text{n}\ \ \ \rightarrow\ \ \ ^7_3\text{Li}^*\ \ \ \rightarrow\ \ \ ^3_1\text{H} + ^4_2\text{He} + 4.78~\text{MeV}
\end{equation}
The experimental signature is a peak at the Q-value of the reaction, i.e.\ 4.78~MeV.
Given the different light yield induced by electrons with respect to alphas and tritons~\cite{ophel},
the thermal neutron peak is recorded at $\sim4.1$~MeV in electron-equivalent energy scale~\cite{birks}.
Since no environmental $\gamma$-ray is commonly present with an energy higher than 2615~keV
(full energy peak of $^{208}$Tl), this assures a very good $\gamma$-$n$\ discrimination.
A major concern for the detection of small thermal neutron fluxes
is the presence of environmental thermal neutrons, as well as
radioactive contaminants of the LiI(Eu) crystal bulk or its surface, emitting $\alpha$\
or $\beta$\ particles in the 3.5-4.5~MeV range.

\begin{figure}
  \centering
  \subfloat[][The LiI(Eu) crystal coupled to the 1'' PMT and mounted on the copper holder.\label{fig:LiI}]{\includegraphics[width=0.48\textwidth]{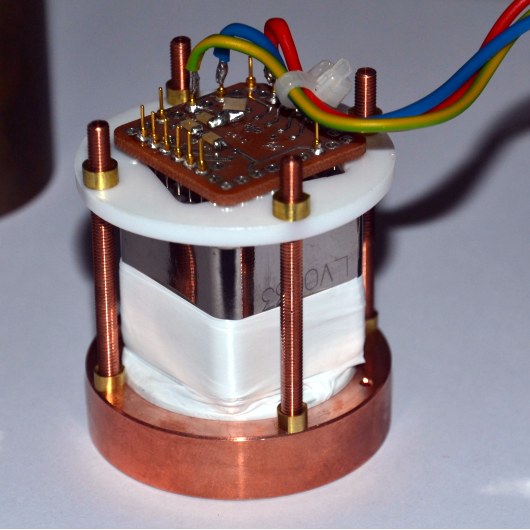}} \quad
  \subfloat[][The detector (in the copper case) in the borated PE shielding,
    together with the 5~cm thick PE moderator and the 2~cm thick Pb absorber.
    The sources were mounted on the holder in front of the Pb brick.\label{fig:cavity}]{\includegraphics[width=0.48\textwidth]{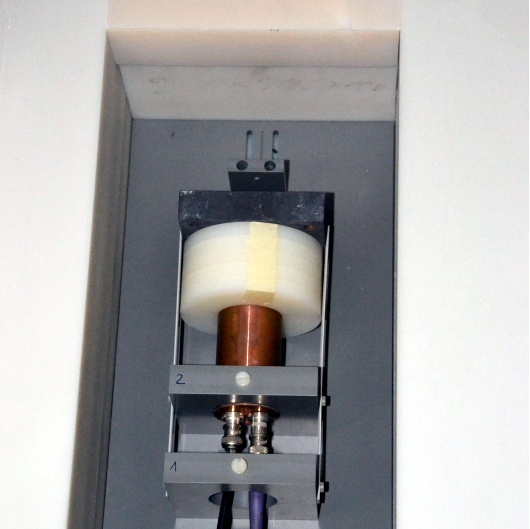}}
  \caption{The low-background LiI(Eu) detector setup.}
\end{figure}

The employed experimental setup consists of a LiI(Eu) cylindrical crystal with 25.4~mm diameter and 3~mm height
enriched to $96\%$\ in $^6$Li provided by Scionix, Bunnik.
The crystal is coupled to a 1'' square R8250 low background photomultiplier tube (PMT) by Hamamatsu Photonics.
In order to minimize the background, the system is encased in a custom made copper holder (Fig.~\ref{fig:LiI}).
The detector is surrounded by 20~cm thick borated polyethylene (PE) panels
acting as a shielding for external neutrons.
A 2~cm thick lead brick and a 5~cm thick PE moderator were present in front of the detector
during all measurements (Fig.~\ref{fig:cavity}).
The detector was designed in the past to measure the neutron fluxes of the \Th\ prototype sources
produced during Gerda Phase I. A detailed description can be found in~\cite{tarka}.

\begin{figure}[]
  \centering
  \def\svgwidth{\textwidth}
  \input{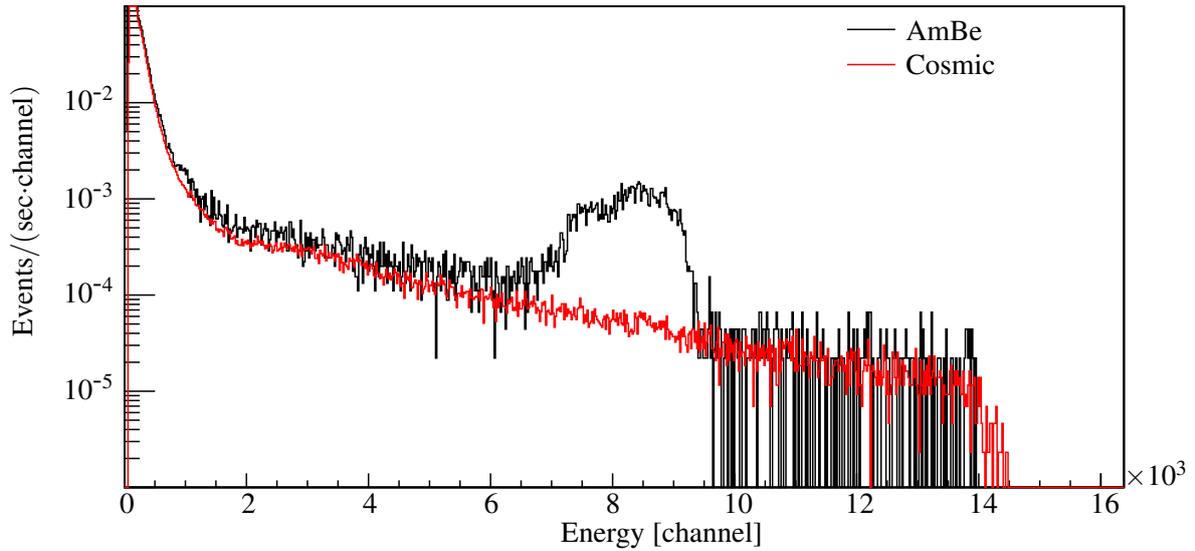}
  \caption{  \label{fig:AmBeSpectrum}
    LiI spectra recorded with and without the AmBe source. The peak due to thermal neutrons is clearly visible
    around channel 8000, corresponding to about 4.1~MeV in electron-equivalent energy scale.
    For these measurements only, the LiI(Eu) detector was operated at the surface laboratory at LNGS.}
\end{figure}

The estimation of the total efficiency $\varepsilon$\ was performed using an AmBe source
with a neutron strength of $160\pm4$~n/s~\cite{weber}.
For this measurement only, the setup was installed at the surface laboratory at LNGS.
The AmBe source was measured for 12.56~h, and a 5~days long background run was taken under the same conditions.
The collected spectra are shown in Fig.~\ref{fig:AmBeSpectrum}.
The analysis was performed via a maximum likelihood analysis in the 7000-9500~bin range,
with $\varepsilon$\ as the parameter of interest:
\begin{equation}\label{eq:LiIEffLikelihood}
\ln{\mathcal{L}} = \ln{ \Biggl( \frac{ \lambda_{s}^{k_{s}} \ \ e^{-\lambda_{s}}}{ k_{s}! } \Biggr) } +
\ln{ \Biggl( \frac{ \lambda_{b}^{k_{b}} \ \ e^{-\lambda_{b}}}{ k_{b}! } \Biggr) } -
\frac{ ( S - \mu_S )^2 }{ 2\cdot\sigma_S^2}
\end{equation}
where $k_s$\ and $k_b$\ are the measured number of counts in the fit range
for the AmBe and the background measurement, respectively,
while $\lambda_s$\ and $\lambda_b$\ are the corresponding expectation values,
and $S$\ is the known AmBe neutron flux.
A Gaussian prior is assigned to $S$, with $\mu_S=160$~n$/$s and $\sigma_S=4$~n$/$s.
The expectation values are defined as:
\begin{eqnarray}
  \lambda_s & = & \Delta t_s \cdot ( \varepsilon \cdot S + B ) \\
  \lambda_b & = & \Delta t_b \cdot B
\end{eqnarray}
where $B$\ is the background index in events$/$s in the fit range
and $\Delta t_s$~($\Delta t_b$) is the live time of the AmBe (background) measurement.
The measured efficiency is:
\begin{equation}
  \varepsilon = \Bigl(5.32^{+0.18}_{-0.15}\pm0.27\Bigr)\cdot10^{-4}
\end{equation}
where the first error is the statistical uncertainty of the fit
and the second is the systematic related to the choice of the energy range.

\begin{figure}[]
  \centering
  \def\svgwidth{\textwidth}
  \input{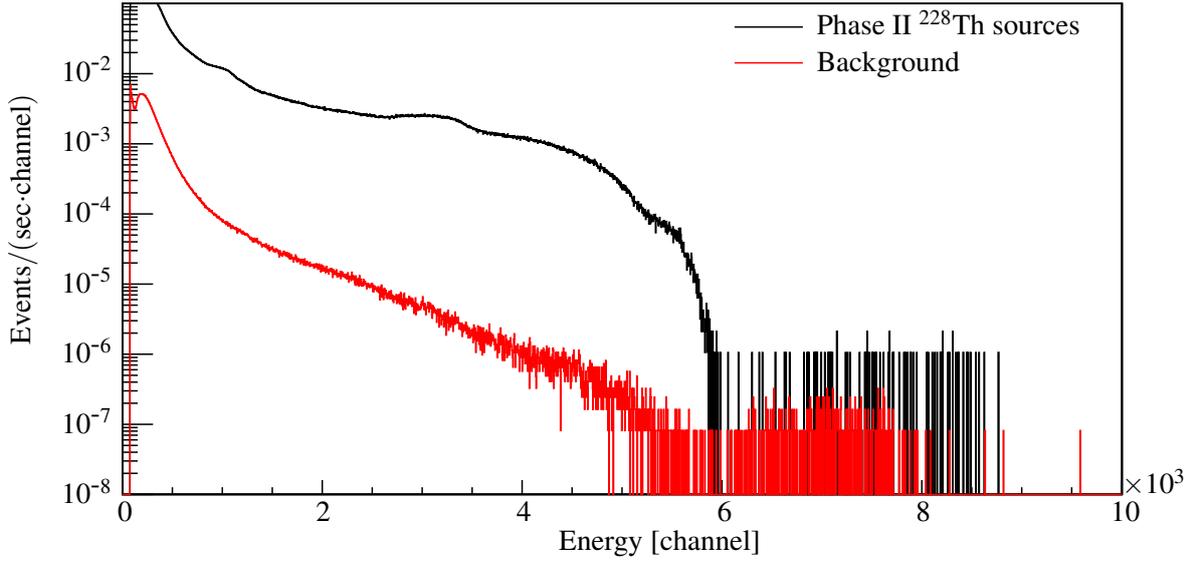}
  \caption{  \label{fig:PhaseIITh228Spectrum}
  Background of the LiI(Eu) detector operated underground at LNGS (red) and spectrum recorded
  with the four \gerda\ Phase II \Th\ sources (black).
  While the $\gamma$\ contribution is only present up to channel $\sim6000$,
  a thermal neutron signal is present around channel $8000$.
  The faint peak in the background spectrum around channel 7000 is likely due to an internal
  $\alpha$~contamination of the crystal.}
\end{figure}

The four \gerda\ Phase II sources were measured together underground for 11 days,
and a background measurement was carried out for 142 days.
Fig.~\ref{fig:PhaseIITh228Spectrum} shows the two recorded spectra.
The estimation of the neutron source strength was performed using the same method as for the efficiency.
The log-likelihood contains a Gaussian prior for the total efficiency
and the total activity $A=130.4\pm5.5$~kBq of the four sources:
\begin{equation}\label{eq:LiIsourcestrength}
\ln{\mathcal{L}} = \ln{ \Biggl( \frac{ \lambda_{s}^{k_{s}} \ \ e^{-\lambda_{s}}}{ k_{s}! } \Biggr) } +
\ln{ \Biggl( \frac{ \lambda_{b}^{k_{b}} \ \ e^{-\lambda_{b}}}{ k_{b}! } \Biggr) } -
\frac{ ( \varepsilon - \mu_\varepsilon )^2 }{ 2\cdot\sigma_\varepsilon^2} -
\frac{ ( A - \mu_A )^2 }{ 2\cdot\sigma_A^2}
\end{equation}
where the subscripts $s$\ and $b$\ refer to the measurement with the sources and the background, respectively.
The expectation values are given by:
\begin{eqnarray}
  \lambda_s & = & \Delta t_s \cdot ( \varepsilon \cdot A \cdot S + B ) \\
  \lambda_b & = & \Delta t_b \cdot B
\end{eqnarray}
The resulting neutron source strength is:
\begin{equation}
  S = \Bigl( 8.2^{+1.7}_{-1.2}\pm1.1 \Bigr) \cdot 10^{-7} \mathrm{n}/(\mathrm{sec}\cdot\mathrm{Bq})
\end{equation}
The first uncertainty is the statistical error from the fit,
while the second is the systematic uncertainty related to the choice of the energy range.
This in particular includes the effect of the background peak
present around channel 7000 (see Fig.~\ref{fig:PhaseIITh228Spectrum}).

The estimated neutron source strength is based on the assumption
that the AmBe and \Th\ emitted neutron spectra are the same.
As mentioned in \cite{tarka,werner}, while the first has a mean energy at $\sim4$~MeV,
gold-encapsulated \Th\ yields a mean neutron energy of 2.58~MeV.
The use of the same total efficiency for the AmBe and \Th\ measurements thus induces a systematic error,
which was computed to be $12\%$\ in~\cite{werner}.
As a result, the neutron source strength for the \gerda\ Phase II calibration sources is:
\begin{equation}
  S = \Bigl( 8.2^{+1.7}_{-1.2} (\text{stat}) \pm1.1 (\text{fit range}) \pm 1.0 (\text{eff}) \Bigr) \cdot 10^{-7} \mathrm{n}/(\mathrm{sec}\cdot\mathrm{Bq})
\end{equation}

For comparison, the same measurement and analysis was performed for a commercial \Th\ source
(i.e., with ceramic substrate) with $19.5\pm2.9$~kBq activity.
The neutron source strength in this case is $7.5^{+2.5}_{-1.3}\cdot10^{-6}$~n$/($sec$\cdot$Bq$)$.
Thus, the use of gold as a support material for \Th\ yields
a reduction in the emitted neutron flux by one order of magnitude.

\subsection{Measurement with a $^{\mathbf{3}}$He counter}\label{subsec:3He}

A second measurement of the neutron source strength was performed with a $^{3}$He counter.
The exploited reaction is:
\begin{equation}
  ^3_2\text{He} + \text{n} \rightarrow ^3_1\text{H} + \text{p} + 746~\text{keV}
\end{equation}
The expected signal induced by thermal neutrons is therefore a peak at 746~keV.
In spite of this low Q-value, $\gamma$\ radiation does not represent a background for the measurement
because of the gaseous state of the detector, with a mean energy deposit for electrons of about $2$~keV/cm.
The discrimination between neutrons and gammas or electrons is thus performed
based on the deposited energy alone.

The $^{3}$He counter is a LND-2531 produced by LND inc., New York.
The tube has an effective diameter of 47.75~mm and a height of 203.2~mm.
It is filled with $^3$He at a pressure of 6078~mbar and is operated at 1950~V.
The counter is inserted in a 6.5~cm thick $4\uppi$\ PE moderator, on which a borehole is present for source insertion.
The distance between the tube and the bottom of the borehole is 1.1~cm.
The system is shielded by a $^{113}$Cd foil and a $4\uppi$\ castle of borated paraffin with 10~cm thickness
for the absorption of environmental neutrons.

The four sources were screened for 17.5~hours, while the background spectrum was acquired for 10~days.
The total detection efficiency, determined using an AmBe source in a previous measurement, is $(7.13\pm0.85)\cdot10^{-2}$~\cite{puccio}.
Following the same analysis procedure used for the LiI(Eu) measurements, the neutron source strength
of the \gerda\ Phase II sources results in:
\begin{equation}
  S = \Bigl( 9.4^{+2.0}_{-0.8}(\text{stat}) \pm0.4 (\text{fit range}) \pm 1.1 (\text{eff}) \Bigr) \cdot 10^{-7} \mathrm{n}/(\mathrm{sec}\cdot\mathrm{Bq})
\end{equation}
where the first uncertainty is statistical, the second is the systematic due to the choice of the energy range,
and the third is the efficiency related systematic.

\begin{figure}[]
  \centering
  \def\svgwidth{\textwidth}
  \input{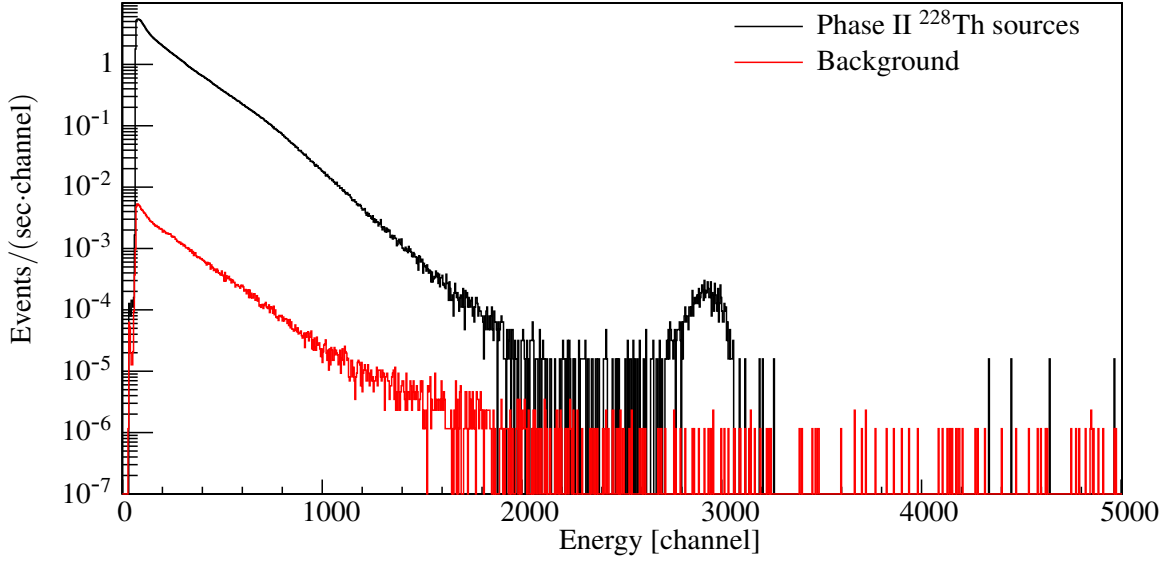}
  \caption{  \label{fig:He3PhaseIITh228}
    Background of the $^3$He counter operated underground at LNGS (red) and spectrum recorded
    with the four \gerda\ Phase II \Th\ sources (black).
    The thermal neutron peak is visible around channel 2900,
    while the broad continuum below channel 2000 is due to $\gamma$\ background.}
\end{figure}

\subsection{Interpretation of results}\label{subsec:interpretation}

The neutron source strengths measured with the LiI(Eu) and the $^3$He detector agree within the errors.
This demonstrates the reliability of the LiI(Eu) as a low-flux neutron detector.
For both measurements, the total uncertainty of $\sim20\%$\ is mostly due to the limited statistics.
A higher precision could be obtained with a longer screening,
but this was not possible due to time constraints related to the preparation for \gerda\ Phase II.

Previous measurements performed with two custom \Th\ sources for \gerda\ Phase I
lead to similar results~\cite{tarka}. The measured neutron source strengths
vary between $7.59\cdot10^{-7}$~n$/($sec$\cdot$Bq$)$ and $1.50\cdot10^{-6}$~n$/($sec$\cdot$Bq$)$,
depending on the source and detector. In this case the uncertainties
were $\sim30\%$, which might explain the spread of the values by up to a factor two.
A more precise measurement was performed in~\cite{werner} on a \Th\ source
with gold substrate and the same encapsulation of the \gerda\ sources.
The measured neutron source strength of $(1.22\pm0.17)\cdot10^{-6}$~n$/($sec$\cdot$Bq$)$
is also in good agreement with the results presented here.

The agreement of independent measurements on the gold-encapsulated \Th\ sources confirms the reliability of the production technique.
The reproducibility of \Th\ sources with neutron emission at the level of $10^{-6}$~n$/($sec$\cdot$Bq$)$
can thus be guaranteed to present and future low background experiments.
In the case of \gerda\ Phase~II, the neutrons emitted by the calibration sources induce a background index at \Qbb\
of $\sim2\cdot10^{-4}$~\eventsper\ prior to the application of PSD and LAr veto.
This is a factor 5 lower than the total expected background
and is sufficient to fulfill the Phase~II requirements.

\section{Leak test for calibration sources in cryogenic systems}\label{sec:leaktest}

A new set of leakage tests intended to check the tightness 
of \Th\ radioactive sources after deployment at cryogenic temperatures
has been developed at the Istituto Nazionale di Metrologia delle Radiazioni Ionizzanti of ENEA (ENEA-INMRI).
It consists of a series of source insertions in acetone and liquid nitrogen,
each one followed by a search for possible leaks.
A blank source, ideally identical to the \gerda\ \Th\ sources but with no radioactive content,
was used for a cross-check.
Each of the four radioactive sources and the blank source 
was enveloped in circular aluminum foil of 24~cm$^2$\ surface, and dipped in acetone (A1) for 1~hour. 
After acetone evaporation, the $\alpha$\ contamination of the aluminum foil was measured
with a Berthold LB~770 $\alpha\mbox{-}\beta$\ low level counter.
The operation was repeated using liquid nitrogen (LN) and then one more time with acetone (A2).
The same procedure, without the second immersion in acetone,
was applied without using any radioactive or blank source for background determination.
While the background measurement is needed to estimate the intrinsic contamination of aluminum,
the comparison between the $\alpha$\ count rates obtained with the real and blank sources allow to 
disentangle between real \Th\ leaks and possible removable surface contamination.
Since the blank and the background count rates are not distinguishable,
they have been considered as belonging to the same population (BB).

\begin{table}
  \caption{Experimental conditions and results for the leak test performed at ENEA-INMRI.}\label{tab:leaktest}
  \centering
  \begin{tabular}{ll}
    \toprule
    Number of BB readings & 108 \\
    Number of A1, LN, A2 readings for each source & 6 \\
    Counting time [s] & 6000 \\
    Mean BB count rate $\bigl[s^{\mbox{-}1}\bigr]$ & 0.00250 \\
    Experimental standard deviation of BB count rate $\bigl[s^{\mbox{-}1}\bigr]$ & 0.00087 \\
    Expected Poisson standard deviation $\bigl[s^{\mbox{-}1}\bigr]$ & 0.00064 \\
    Decision threshold for the net individual count rate $\bigl[s^{\mbox{-}1}\bigr]$ & 0.00202 \\
    \bottomrule
  \end{tabular}
\end{table}

The distribution of A1, LN and A2 count rates for each of the four sources was compared with that of BB.
The test conditions and the results of the analysis are reported in Tab.~\ref{tab:leaktest}.
Following \cite{iso11929}, the decision threshold at $95\%$ confidence level
was calculated based on the standard deviation of the BB readings.
The experimental standard deviation of the BB includes components of variability due to counting statistics,
long term system stability and homogeneity of used materials.
If the sample reading of the sources exceeds the decision threshold, a \Th\ leak has been detected.

Out of all the 24 LN readings only $2~(8.3\%)$\ exceed the decision threshold, while they are in average lower than the A1 and A2 readings.
In addition, the A1 counting rates are always slightly higher than those of A2.
This means that the higher LN counting rates is not induced by a \Th\ leak,
but from a surface contamination of the source capsules,
which is removed by the two insertions in acetone.

The activity removed from each of the sources after the A1, LN and A2 immersions
is reported in Tab.~\ref{tab:leakactivities}. The efficiency of the counter was measured
with a standard $^{241}$Am source with known activity (555.64~Bq) and is 0.3968.
For all sources, no leak is detected after the insertion in LN and the second insertion in acetone.
Hence the sources are suited to be used in \gerda\ Phase~II.

\begin{table}
  \caption{Activity found after each source treatment,
    with reported $2~\sigma$\ uncertainties.}\label{tab:leakactivities}
  \centering
  \begin{tabular}{lccc}
    \toprule
    Source & A1 [mBq] & LN [mBq] & A2 [mBq] \\
    \midrule
    AD9854 & $7.5\pm3.5$  & $1.0\pm6.3$         & $4.4\pm3.8$ \\
    AD9855 & $4.2\pm4.6$  & $1.0\pm6.1$         & $4.0\pm3.0$ \\
    AD9856 & $1.0\pm2.9$ & $\mbox{-}0.8\pm2.7$ & $\mbox{-}0.2\pm3.6$ \\
    AD9857 & $3.2\pm3.8$  & $2.6\pm6.4$         & $0.8\pm3.9$ \\
    \bottomrule
  \end{tabular}
\end{table}

\section{Summary}\label{sec:summary}

Four low neutron emission \Th\ source were produced for the calibration of the \gerda\ Phase II experiment.
With a threshold of 9.94~MeV for $(\alpha,n)$\ reactions,
the deposition of the radioactive material on gold allows to minimize the parasitic neutron flux
and allows to reduce the neutron source strength by one order of magnitude.
The emitted neutron flux was measured with a low background LiI(Eu) detector and a $^{3}$He counter,
both located underground at LNGS.
The neutron source strengths measured with the LiI(Eu) and the $^3$He detectors are
$8.2\pm2.3$~n$/($sec$\cdot$Bq$)$\ and $9.4\pm2.3$~n$/($sec$\cdot$Bq$)$, respectively.

The $\gamma$\ activity of the each source was measured with the Gator facility at LNGS.
The high signal-to-background ratio, together with a detailed knowledge of the detector geometry,
allowed to obtain a very good agreement between measured and simulated spectra, with a total uncertainty of $\sim4\%$.

A new, dedicated procedure was developed for the investigation of the source capsule tightness
after the deployment in cryogenic environment,
with a sensitivity to leaks of $\lesssim10$~mBq activity.
The application of this technique to the \gerda\ Phase II calibration sources
excluded any leaks and guarantees the tightness of the source capsules after the insertion in \gerda.

\acknowledgments

The work was financially supported by the University of Zurich, the Swiss National
Science Foundation (SNF) grants No. 200020-149256 and No. 20AS21-136660, and by the INT Invisibles
(Marie Curie Actions, PITN- GA-2011-289442).
The authors sincerely thank the Institut f\"ur Kernchemie in Mainz, Germany,
the ENEA Istituto Nazionale di Metrologia delle Radiazioni Ionizzanti (INMRI), Italy,
and the Laboratori Nazionali del Gran Sasso (LNGS) of INFN, Italy,
for the fruitful scientific cooperation.
We also thank M.~Laubenstein, A.~Giampaoli, A.~D.~Ferella and C.~Macolino
for their help during the measurements performed at LNGS.

\end{document}